\documentclass[aps,prb,reprint]{revtex4-1}

\usepackage{graphicx}
\usepackage{dcolumn}
\usepackage{bm}

\usepackage[utf8]{inputenc}
\usepackage[T1]{fontenc}
\usepackage{mathptmx}

\begin{document}

\title[Jump of tunneling magnetoresistance in magnetic nanocontacts with mismatched cross section]{Jump of tunneling magnetoresistance in magnetic nanocontacts with mismatched cross section}

\author{V.L. Katkov}
 \altaffiliation{Bogoliubov Laboratory of Theoretical Physics, Joint Institute for Nuclear Research,
141980 Dubna, Moscow region, Russia.}
\author{V.A. Osipov}
\affiliation{Bogoliubov Laboratory of Theoretical Physics, Joint Institute for Nuclear Research,
141980 Dubna, Moscow region, Russia.
}

\date{\today}

\begin{abstract}
We have studied the influence of the transverse size of a magnetic tunnel nanojunction on the magnitude of the magnetoresistance. During modeling, the size of the right contact was fixed, while the size of the left one gradually changed until they coincided.  We found a sharp drop in the tunneling magnetoresistance (TMR) in nanocontacts with  mismatched cross section. This can be explained by the peculiarities of the spatial distribution of the electron density, which is different for majority and minority-spin states. The discovered effect must be taken into account in the design of TMR-based nanodevices.
\end{abstract}

\maketitle

\section{Introduction}

Magnetoelectronic devices, both existing and proposed, are based on spin-dependent electron transport. For example, modern read heads for high-density data storage devices operate using the ``giant'' magnetoresistance (GMR) effect, a spin-dependent effect observed over three decades ago in magnetic multilayers~\cite{Baibich1988, Binasch1989}. This discovery was awarded the Nobel Prize in 2007. A historical overview summarizing the related research by that time is presented in Ref.~\onlinecite{Fullerton2007}.

At the same time, not only GMR, but also  ballistic magnetoresistance (BMR) and tunnel magnetoresistance (TMR) phenomenons were found in magnetoelectronic devices (for example, in nanoscale magnetic contacts). The observed effects depend on the choice of materials as well as the design of the contact. The BMR concept implies a two-component ferromagnet/barrier/ferromagnet as well as a three-component ferromagnet/conductor/ferromagnet junction, in the absence of diffusion transport both in the contacts and in the channel. Usually, the term TMR refers only to the ferromagnet/barrier/ferromagnet  configuration (see Ref.~\onlinecite{Fullerton2007} and the references therein). However, often these terms are used interchangeably, as discussed in Ref.~\onlinecite{Garcia2000b}.

The TMR effect for bulk materials was discovered and described in 1975~\cite{JULLIERE1975}. However, more recent research by Garcia et al.~\cite{Garcia1999, Tatara1999, Garcia2000} has given new impetus to the study of electron transport in ferromagnets. The authors found evidence of BMR in {\it nanoscale} magnetic contacts. For this purpose, they managed to build a nanocontact between two thin nickel wires, where each is controlled by its own magnet. Experiments have shown that the mean free path of the electron is larger than the polarized area size in each of the contacts and, therefore, the transport can be called ballistic. The magnitude of BMR was found to be 280 \% for Ni-Ni~\cite{Garcia1999} and 200 \% for Co-Co~\cite{Tatara1999} nanoscale contacts at room temperature. An order of magnitude smaller value of BMR (about 30 \%)  was observed in Fe-Fe contacts\cite{Garcia2000}.

Some authors have studied the effects associated with the geometric parameters of the domain wall and its dynamics~\cite{Versluijs2001, Yamaguchi2004, Noh2011}, the geometry and electronic structure of the contacts~\cite{Jacob2005}, as well as the type of the contact separating area~\cite {Zhuravlev2001, Zhuravlev2003}.
 In recent works~\cite{Useinov2016, USEINOV2018}, the behavior of TMR in tunnel junctions with embedded nanoparticles was investigated. The maximum value of TMR in such structures was found to lies close to 300 \%.

In this paper, our attention will be focused on the TMR effects associated exclusively with the transverse size of the contacts. We use the method proposed in Refs.~\onlinecite{Mathon1997, Mathon1997B, Mathon2001} and based on the application of nonequilibrium Green's functions in the framework of the {\it tight-binding} approximation. In the simulation, we consider contacts having a cubic lattice. In this case, the edge Green's functions are known analytically, which makes it possible to investigate mesoscopic structures of a sufficiently large cross section, on the order of ten nanometers, and this corresponds to the typical experimental values. A similar problem for more advanced methods based, for example, on computations within the framework of the density functional theory (DFT), is extremely expensive in terms of computational resources and usually allows describing structures with a cross section of about 1 nm (see, for example, Refs.~\onlinecite{Sabirianov2005, Hope2008}). Another research method widely used for this kind of problems, the free electron model, is not adapted to take into account the effects associated with the spatial distribution of electrons and edge effects. As we will show, this factor is significant when considering contacts with a nanometer-sized cross section.

\section{Model}

The system under study consists of two contacts of small cross-section, the magnetization in which can be controlled, achieving a parallel and antiparallel configuration of spins in each of them.
In this work, we use a model similar to that of Matton~\cite{Mathon1997, Mathon2001}. The essence of the model is to use the standard method of nonequilibrium Green's functions to calculate the conductivity for each magnetic configuration. The {\it magnetoresistance ratio} is defined as
\begin{equation}
    TMR =\frac{\Gamma_{P}^\uparrow +  \Gamma_{P}^\downarrow - (\Gamma_{AP}^\uparrow + \Gamma_{AP}^\downarrow)}{\Gamma_{AP}^\uparrow + \Gamma_{AP}^\downarrow} \times 100 \% ,
    \label{rTMR}
\end{equation}
where $\Gamma_{C}^\sigma$, is the conductivity, $C=(P, AP)$ is the type of configuration (parallel and antiparallel), $\sigma=(\uparrow,\downarrow)$ is the spin channel. In Ref.~\onlinecite{Mathon1997}, the case of a macroscopic (infinitely wide) contact with a cubic lattice was considered for which the conductivity is expressed in terms of the edge Green's functions in a usual way:
\begin{equation}
    \label{eq:1}
   \Gamma(g_L, g_R) = \frac{4e^2}{h}\sum\limits_{\bf{k}_{||}} t_{gap}^2 \frac{\text{Im } g_L^-(E_F,{\bf{k}}_{||}) \text{ Im } g_R^-(E_F,{\bf{k}}_{||})}{|1-t_{gap}^2g_L^-(E_F,{\bf{k}}_{||})g_R^-(E_F,{\bf{k}}_{||})|^2}.
\end{equation}
Here $g_{S}^{+}(E_F, {\bf{k}}_{||})$ is the retarded surface (edge) Green's function of the left and right contacts ($S = L, R$) taken at the Fermi level, $g_{S}^{-} = (g_{S}^{+})^*$, $t_{gap}$ is the hopping integral between the nearest atoms of neighboring contacts, summation is performed over all values of the momentum in the direction perpendicular to direction of transport. The majority and minority spin states are taken into account by locating the center of the bands at the points $\varepsilon_{FM}\mp\Delta/2$, respectively, where $\varepsilon_{FM}$ is the spin-independent on-site potential in
the ferromagnet and $\Delta$ is the exchange splitting of the bands. The Green's function for each state is defined as $g_S^{\Uparrow\Downarrow}(E_F) = g_S(\varepsilon_{FM}\mp\Delta/2)$, where up and down arrows indicate belonging to the majority  and minority spin band, respectively. As a result, the conductivities for each spin channel ($\sigma=\uparrow\downarrow$) and two configurations of contacts ($P$ and $AP$) have the following form: $\Gamma^\uparrow_P=\Gamma(g_L^\Uparrow, g_R^\Uparrow)$, $\Gamma^\downarrow_P = \Gamma(g_L^\Downarrow, g_R^\Downarrow)$, $\Gamma^\uparrow_{AP} = \Gamma(g_L^\Uparrow, g_R^\Downarrow)$, $\Gamma^\downarrow_{AP}=\Gamma(g_L^\Downarrow, g_R^\Uparrow)$. Hereinafter, the conductivity of the majority spin channel of the left contact is denoted by $\uparrow$.

Comparison of $TMR$ with the results obtained in the tight-binding approximation for cobalt showed that the cubic lattice model mimics the cobalt when the ferromagnetic parameters  are chosen to be $\varepsilon_{FM} = - 5.1 t$ and $\Delta = 1.0 t$, where $t$ is the hopping integral in a square lattice (the Fermi level is taken as zero)~\cite{Mathon1997}. We will use these ferromagnetic parameters in our calculations.

\begin{figure}[!ht]
\centering
			\includegraphics[width=0.32\textwidth]{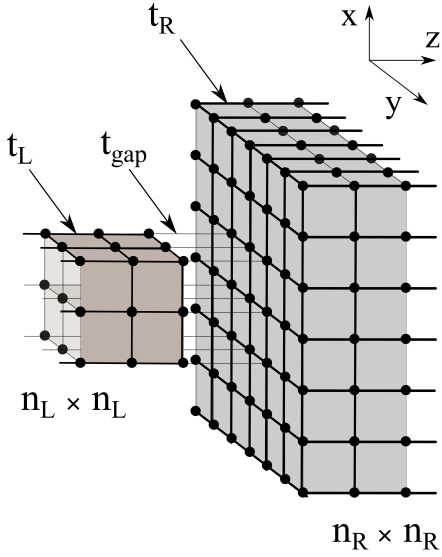}
		\caption{Schematic illustration of a tunnel nanojunction with mismatched cross section.}
		\label{fig:1}
\end{figure}

Let us consider magnetic contacts having different finite transverse size (see Fig.~\ref{fig:1}). In this case, the above described approach requires a certain modification. This is achieved by using the generalized expression~\cite{Katkov2017} for the conductivity instead of~(\ref{eq:1}) 
\begin{eqnarray}
 \Gamma &=& \frac{e^2}{h}{\text{Tr}}[{\bf A}_L({\bf{1}}-{\bf t}_{gap}^{\dagger}{\bf g}_R^-{\bf t}_{gap}{\bf g}_L^-)^{-1}{\bf t}_{gap}^{\dagger} \nonumber \\  
&  \times & {\bf A}_R{\bf t}_{gap}({\bf{1}}-{\bf g}_L^+{\bf t}_{gap}^{\dagger}{\bf g}_R^+{\bf t}_{gap})^{-1}], 
\label{eq:2}
\end{eqnarray}
where ${\bf A}_{S}=i({\bf g}_{S}^+ - {\bf g}_{S}^-)$ is the spectral density in matrix form, ${\bf g}_S^{\pm}$ are retarded and advanced surface (edge) matrix Green's functions with dimension $n_S\times n_S$, ${\bf t}_{gap}$ is a matrix describing the interaction of contacts (with dimension $n_L \times n_R$), ${\bf 1}$ is the identity matrix.

In the case of a cubic lattice, all elements of the matrix ${\bf g}_S^{+}$ can be found analytically and have the following form~\cite{Berthod2011}:
\begin{eqnarray} \nonumber
         g_{S, ij}^{+}(\varepsilon) &=&  \frac{4/t_S}{(n_S+1)^2}\sum\limits_{k_x, k_y} \sin(k_x x_i) \sin(k_x x_j)\sin(k_y y_i) \\ 
         &\times& \sin(k_y y_j)L\left(\frac{\varepsilon - 2 t_S (\cos k_x + \cos k_y)+ i 0^+}{2 t_S}\right),\\
         k_{x (y)} &=& n_{x (y)} \pi / (n_S + 1),~~ n_{x (y)} = 1, 2, \dots n_S,\\
         L(z) &=& z - \sqrt{z+1} \sqrt{z-1},
         \label{GF_L}
\end{eqnarray}
where $(x_{i}, y_{i})$ and ($x_j, y_{j})$ are the coordinates of atoms with indices $i$ and $j$ at the edge of the considered contact.
Hereinafter, the lattice parameter is taken as unity ($a=1$).

DFT calculations of the densities of states of bulk cobalt and ultrathin nanowires (only a few tens of atoms per unit cell) show that parameters responsible for a position of the Fermi level and a relative displacement of the bands practically do not  change when passing from a bulk crystal to a wire~\cite{Hope2008}. It was assumed that the material on the left and right is the same, $t_L = t_R = t$,  and, respectively, $\varepsilon_{FM}$ and $\Delta$ are identical for both contacts. A well-defined model with an abrupt domain wall was considered when $t_{gap}=t$. In the case of weaker coupling between contacts, the behavior turned out to be similar to that of infinitely wide contacts described in~\cite{Mathon1997, Mathon2001}. Its main feature is the rapid approach the $TMR$ to the saturation value, which then changes slightly with  decreasing coupling parameter $t_{gap}$.

\begin{figure}[!ht]
\centering
			\includegraphics[width=0.45\textwidth]{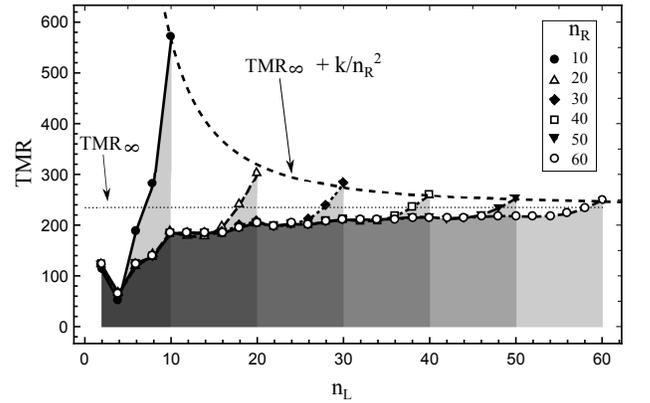}
		\caption{$TMR$ for $n_R$ varying in the range from 10 to 60 with a step of 10 and $n_L$ varying in the range from 2 to $n_R$ with a step of 2. $TMR_\infty = 234.3$ is the calculated value for infinitely wide contacts. $TMR$ for $n_L=n_R$ is well described by the dependence $TMR_\infty + k/n_R^2$, where $k = 33277.5$ is a fitted parameter.}
		\label{fig:2}
\end{figure}

\section{Results and discussion}

In the calculations, the transverse size of the right contact $n_R$ was fixed in the range from 10 to 60 with a step of 10. The maximum value $n_R = 60$ corresponds to a wire cross section of several tens of nanometers. In this case, the lateral size of the left contact $n_L$ varied from 2 up to $n_R$ with a step of 2. The obtained dependencies of the $TMR$ are shown in Fig.~\ref{fig:2}. The value of $TMR_\infty$ for infinitely wide junction calculated with (\ref{eq:1}) is shown as a horizontal line. We found that the behavior of $TMR_e$ on $n$ for contacts of the same size is well described by the inverse quadratic dependence $TMR_e = TMR_\infty + k/n^2$, which has an obvious limit $TMR_\infty$ for $n\to\infty$.

Fig.~\ref{fig:2} shows an increase in the magnetoresistance ratio  with decreasing contact sizes. This effect was discovered experimentally and is usually explained by the fact that the density of electronic states changes with a change in the transverse size of polarized contacts~\cite{Garcia1999, Tatara1999}.

The most significant result of our research is the finding of a sharp drop in $TMR$ even at a small difference in the cross sections of the contacts, which occurs for all considered sizes. As we show below, this effect has a clear physical interpretation. 

First of all, it should be noted that, according to our analysis, the main contribution to $TMR$ as a function of 
$n_L $ comes from $\Gamma^\downarrow_P$ and $\Gamma^\downarrow_ {AP}$, which are included in the numerator and denominator of the definition (\ref{eq:1}), respectively. Recall that $\Gamma^\downarrow_P$ is the conductivity for the minority spin channel in the absence of a domain wall, while $\Gamma^\downarrow_ {AP}$ is the conductivity for the minority channel in its presence. We found that $\Gamma^\downarrow_P$ grows with an increase in the size of the left contact, as would be expected. At the same time, $\Gamma^\downarrow_{AP}$ stops growing after a certain value of $n_L$, what is a common trend for all considered values of $n_R$. 

\begin{figure}[!ht]
\centering
			\includegraphics[width=0.45\textwidth]{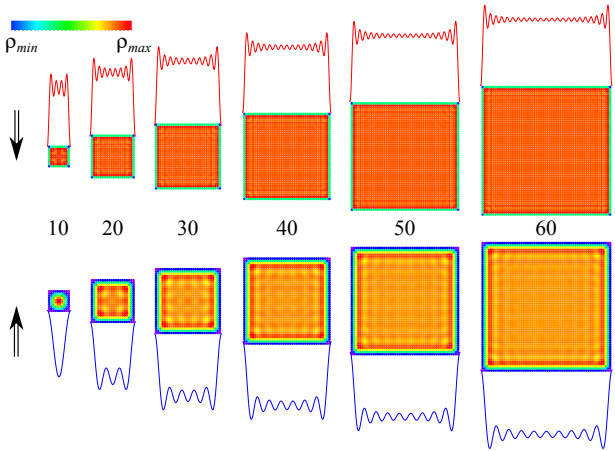}
		\caption{Local density of electronic states at the edge of contacts of different transverse sizes ($n_S = 10\div 60$) for two spin bands. The curves show the results of calculating the one-dimensional density by Eqs.~(\ref{eq:9}) and (\ref{eq:11}). It can be seen that for the majority spin band electrons are always localized closer to the center of the wire, which is determined by the maximum Fermi momentum (or the minimum electron wavelength at the Fermi level).}
		\label{fig:3}
\end{figure}

The observed behavior of $TMR$ can be clarified by visualizing the density of electronic states at the Fermi level for atoms at the edges of contacts for each of the spin bands. The local density of edge states (which is proportional to the diagonal elements of the spectral density matrix ${\bf A}_{S}=i({\bf g}_{S}^+ -{\bf g}_{S}^ -)$) is shown in Fig.~\ref{fig:3}. It is seen that in the $\Downarrow$ state the electrons are  uniformly distributed over the entire contact area, while in the $\Uparrow$ state the significant electron depletion always takes place in the boundary region. This is explained by the fact that the $\Uparrow$ states are characterized by a smaller maximum value of the Fermi momentum $k_F$ in the transverse direction and, accordingly, a longer wavelength, which leads to an increase in the electron-depleted region near the edge.

For illustration, consider the electron density along one of the transverse directions, for example, along the $x$ direction. For a three-dimensional cubic lattice at the Fermi level (taken as zero), the following equality holds:
\begin{equation}\label{eq:7}
2 t (\cos k_{Fx}^{\Uparrow \Downarrow}  +  \cos k_{Fy}^{\Uparrow \Downarrow} + \cos k_{Fz}^{\Uparrow \Downarrow})+ \varepsilon_{FM}\mp \Delta/2 = 0.     
\end{equation}
The maximum value of the wavevector $k_{Fx}^{max}$ can be obtained from ($\ref{eq:7}$) provided $k_{Fy}=k_{Fz}=0$:
\begin{equation}\label{eq:8}
    (k_{Fx}^{max})^{\Uparrow\Downarrow} = \arccos\left[\pm \Delta/(4t) -\varepsilon_{FM}/(2t) - 2\right].
\end{equation}
By definition, the local density of states along the chosen direction is
\begin{equation}\label{eq:9}
    \rho(x) \propto \sum\limits_n |\psi_n(x)|^2 = \sum\limits_{k_x}^{k_x^{max}} \sin (k_x x)^2, 
\end{equation}
where $k_x = i \pi/(n_S+1)$, $i = 1,2\ldots i^{max}.$
Moreover, the maximum term in this sum should be found from the condition
\begin{equation}\label{eq:11}
   \frac{\pi (i^{max})^{\Uparrow\Downarrow} }{n_S+1} <  (k_{Fx}^{max})^{\Uparrow\Downarrow}. 
\end{equation}
Smaller $k_x$ correspond to non-minimal values of $k_y$ and $k_z$, such that the equality (\ref{eq:7}) holds.
The graphs corresponding to the selected ferromagnetic parameters for $\rho(x)$, obtained using Eq.~(\ref{eq:9}), are shown in Fig.~\ref{fig:3} for $k_{Fx}^{max\Uparrow}$ and $k_{Fx}^{max\Downarrow}$.

Let the left contact be in the state  $\Downarrow$ and the right contact in the state $\Uparrow$. Then, with an increase in the transverse size of the left contact $n_L$, the conductivity will grow up to the moment when the left contact coincides in size with a region of the right contact in which electronic states are still present. Further increase in $n_L$ will not cause any increase in conductivity. In our case, this behavior leads to saturation of $\Gamma^\downarrow_{AP}$ at $n_L \geq n_R - 4$, which, in turn, determines the characteristic growth of $TMR$ after $n_L$ reaches this value (see Fig.~\ref{fig:2}).

\section{Conclusion}
Our study showed that the depletion of the $\Uparrow$ electronic states near the boundaries of the contact cross-section is the reason for the marked increase in $TMR$. This effect is most pronounced for nano-sized contacts. Namely, we found that the growth of $TMR$ with decreasing width of the square contact is well described by the $n^{- 2}$ law. Despite the fact that the calculations were performed within the mono-orbital model, the obtained results are obviously of a general nature and are applicable to a wide range of ferromagnetic materials with different lattice geometries. Indeed, the essence of the effect is the difference of maximum achievable wavelengths for electrons located at the Fermi level in the $\Uparrow$ and $\Downarrow$ bands which leads to the emergence of electron-depleted region at the edge of the contact cross-section. We found that the more the electron wavelengths differ in different states of polarization, the more noticeable the manifestation of this effect.
For the typical ferromagnetic metals (cobalt, nickel, iron), this difference is quite large, since light electrons from the $sp$-band are responsible for the $\Uparrow$ state, while the $\Downarrow $state is formed mainly by heavy particles from the $d$-band.

Theoretical and experimental estimates of the degree of spin polarization (DSP) $\zeta=(DOS_{F}^{\Downarrow} - DOS_{F}^{\Uparrow})/(DOS_F^\Downarrow + DOS_F^\Uparrow)$ for bulk Co, Ni and Fe are $0.83$, $0.87$ and $0.5$, respectively~\cite{Tatara1999, Garcia2000}. For nanosized {\it hcp} cobalt wires of different thicknesses, this characteristic takes values in the range of ($0.43 \div 0.92$) \cite{Hope2008}. Given that the Fermi momentum  $k_F\propto\sqrt{DOS_F}$, the ratio of wavelengths $\lambda^\Uparrow_F/\lambda^\Downarrow_F =\sqrt{(1 + \zeta)/{(1-\zeta)}} $ lies in the range of ($1.5\div 5$), which is in good agreement with our estimate for the chosen ferromagnetic parameters. Indeed, using (\ref{eq:8}) we get $\lambda^{max\Uparrow}_{F}/\lambda^{max\Downarrow}_{F} = 2.2$.

We would like to emphasize once again that the discovered transport effect associated with the spatial localization of electrons is difficult to reproduce using {\it ab initio} calculations. The point is that the contacts must have a sufficiently large transverse size (about a hundred atoms in the cross section) to accommodate at least several wavelengths of electron density, which makes {\it ab initio} calculations extremely costly.

It should be noted that the discovered phenomena may have important practical consequences. The first is that identical contacts have a significantly higher value of $ TMR $, and even a small mismatching in transverse sizes can lead to a noticeable drop in the magnitude of the magnetoresistance. Thus, in the production of real nanodevices with the highest magnetoresistance ratio, it is necessary to achieve maximum coincidence of the cross sections of ferromagnetic contacts.
On the other hand, this effect can be considered as a tool for precision scanning of nanowire thickness uniformity.
For this purpose, one has to move a domain wall along the wire what can be done in different ways~\cite{Yamaguchi2004, Noh2011}. According to our results, noticeable jumps in $TMR$ should be observed in the area of heterogeneity (thickening or thinning).

\end{document}